\documentclass[12pt]{article}
\begin{document}
\title{Fits to data on polarised structure functions
and spin asymmetries with power law corrections}
\author{ Jan
Bartelski\\ Institute of Theoretical Physics, Warsaw University,\\
Ho$\dot{z}$a 69, 00-681 Warsaw, Poland. \\ \\ \and Stanis\l aw
Tatur
\\ Nicolaus Copernicus Astronomical Center,\\ Polish Academy of
Sciences,\\ Bartycka 18, 00-716 Warsaw, Poland. \\ }
\date{}
\maketitle
\vspace{1cm}
\begin{abstract}
\noindent We have compared polarized parton densities determined
in the NLO QCD fits to polarized structure functions and spin
asymmetries. We consider models of such distributions based on
MRST 99 and MRST 2001 fits to non-polarized data. Simple power law
corrections corresponding to higher twists are taken into account
and their importance is analyzed. The role of positivity
conditions for parton densities and their influence on the values
of $\chi^2$ is discussed.
\end{abstract}
\newpage
In \cite{BTpr} we got parton densities using NLO QCD
fit from the data on polarized structure function $g_{1}$ and spin
asymmetries.
We have used in our fits functional form for parton densities
corresponding to unpolarized distributions obtained in the Martin, Roberts, Stirling and Thorne
(MRST 98) fit \cite{MRSTnew}. The results for integrated parton densities were very
similar in both cases: for ones gotten from the polarized structure functions
and spin asymmetries. In
that comparison we have not taken into account power law
corrections connected with higher
twists contributions. It has been pointed out \cite{Leader} that these
corrections can play a role in the determination of parton densities. It is an aim of the
present paper to consider the influence of the power law
corrections in the different methods of extracting polarized
parton distributions from fits to the experimental data.

As before we will
use all available experimental data on polarized structure
functions and corresponding spin asymmetries
\cite{exp,expEMC,expSMC,expE142,expE143,expHermes,expE155}. Functional form
of our polarized parton densities  will be taken from corresponding
 unpolarized parton densities from the fits made by Martin, Roberts, Stirling and Thorne
(called MRST 99 \cite{MRST99} and MRST 2001 \cite{MRST01}). We will use the same
method of extracting such densities as in our previous
papers \cite{BT,BTn}, where the experimental data for spin
asymmetries were used to obtain polarized parton densities. Many
experimental groups gave also experimental results for polarized
structure functions $g_{1}$ measured for proton, neutron and
deuteron.
The asymptotic
behaviour of our polarized parton distributions is determined (up
to the condition that the corresponding spin densities are
integrable) by the fit to unpolarized data. Experiments on
unpolarized targets provide information on the spin averaged quark
and gluon densities $q(x,Q^2)$ and $G(x,Q^2)$ inside the nucleon.

Our quark distributions (at $ Q^{2}=1\,{\rm GeV^{2}}$) are
parametrized as follows:
\begin{equation}
\Delta q_{i}(x) = A_i x^{\lambda_i} (1-x)^{\eta_i} (1+\epsilon_i
\sqrt{x}+\mu_i x).
\end{equation}
The similar form one has for
gluon density:
\begin{equation}
\Delta G(x) = B_G x^{\lambda_G} (1-x)^{\eta_G} (1+\epsilon_G
\sqrt{x}+\mu_G x).
\end{equation}
The values of constants $\lambda_i$ and $\eta_i$ are given in \cite{MRST99,MRST01}.

\noindent We have for total quark distributions:
\begin{eqnarray}
\Delta u& = &\Delta u_{v}+2 \Delta \overline{u}, \nonumber \\
\Delta d& = &\Delta d_{v}+2 \Delta \overline{d},\\ \Delta s &=&2
\Delta \overline{s}, \nonumber
\end{eqnarray}
and for axial charges:
\begin{eqnarray}
\Delta \Sigma &= &\Delta u+\Delta d+\Delta s, \nonumber \\ a_{8}
&=& \Delta u+\Delta d-2\Delta s, \\ a_{3} &\equiv& g_A= \Delta
u-\Delta d. \nonumber
\end{eqnarray}

In order to determine the unknown parameters in the expressions
for  polarized quark and gluon distributions we calculate the spin
asymmetries (starting from initial value $Q^2$ = 1 $\mbox{GeV}^2$)
for measured values of $Q^2$ and make a fit to the experimental
data on spin asymmetries for proton, neutron and deuteron targets.
The spin asymmetry $A_1(x,Q^2)$ can be expressed via the polarized
structure function $g_1(x,Q^2)$ as:
\begin{equation}
A_1(x,Q^2)\cong \frac{(1+\gamma^{2}) g_{1}(x,Q^2)}{ F_1(x,Q^2)}=
\frac{ g_{1}(x,Q^2)}{ F_2(x,Q^2)}[2x(1+ R(x,Q^2))],
\end{equation}
(In the case of SLAC data \cite{expE143,expE155} $g_{1}/F_{1}$ values were given
experimentally)
\noindent where $R = [F_2(1+\gamma^{2})-2xF_1]/
2xF_1$, whereas $\gamma =2Mx/Q$ ($M$ stands for proton mass). We
will take the  value of $R$ from the \cite{whitn}
(of course formula for $R$ determined experimentally already includes
corrections from higher twists). In calculating
$g_{1}(x,Q^2)$ and $F_{2}(x,Q^2)$ in the next to leading order we
use procedure described in \cite{ewol2,BT, BTpr} (power law corrections
are not included). Having calculated the asymmetries according to
eq.(5) for the value of $Q^2$ obtained in experiments we can make
a fit to asymmetries on proton, neutron and deuteron targets.

As usual in our previous papers the value of $a_3$ is not constrained
in such fit. We will also do not fix $a_{8}$ but we put a
constraint on its value. Simply we will add it as an extra
experimental point (from hyperon decays one has $a_8 = 0.58 \pm
0.03$, but we enhance an error to 3$\sigma$, i.e. to 0.1). It is also
possible to use experimentally determined $F_{2}(x,Q^2)$, for
example by SMC group from CERN, where the power law corrections are
included in the fit \cite{SMCth}. It is also possible to use directly
experimental results for polarized structure functions $g_{1}$
given by many experimental groups (the problem is that $g_{1}$
could be determined in a different way in different experiments).
We will try to use these three methods of obtaining polarized
quark and gluon densities taking into account very simple
($h/Q^2$) power law (with no $x$ dependence) correction to the NLO QCD expression for the
polarized structure function $g_{1}(x,Q^2)$ ($g_{1}(x,Q^2) \rightarrow g_{1}(x,Q^2)+h/Q^2$).
Simple $x$ dependence of coefficient $h(x)$, as well as terrms of order $1/Q^4$, give negligible corrections to $\chi^2$.

\vspace{0.5cm}
\hspace{5cm} Table 1

\vspace{0.2cm}
\hspace{0.5cm}
\begin{tabular}{||c|c|c|c|c|c|c||}
\hline \hline $fit$&$\Delta u$&$\Delta d$&$\Delta s$&$\Delta
\Sigma$&$g_A$&$\Delta G$\\ \hline \hline $A99$&$0.74
$&$-0.48$&$-0.16 $&$0.10$&$1.22$&$0.47$ \\ \hline $B99$&$0.73
$&$-0.49$&$-0.17 $&$0.08$&$1.22$&$0.49$ \\ \hline $C99$&$0.78
$&$-0.48$&$-0.14 $&$0.16$&$1.26$&$-0.43$ \\ \hline \hline
$$&$h_{1p}$&$h_{1n}$&$\chi^2$&$\chi^2_0$&$\chi^2_{(np)}$&$\chi^2_{0(np)}$\\
\hline \hline $A99$&$0.04
$&$0.02$&$165.04$&$177.58$&$153.83$&$156.45$ \\ \hline $B99$&$0.03
$&$0.03$&$159.35$&$165.94$&$152.01$&$154.02$ \\ \hline
$C99$&$0.0008 $&$0.008$&$150.77$&$150.86$&$149.34$&$149.58$ \\
\hline \hline
\end{tabular}
\vspace {0.5cm}

In the Table 1 we present our results for integrated parton
densities with corresponding $\chi^{2}$ for the fit (subscript $0$ corresponds
to the fit with no higher twists corrections, whereas $(np)$ stands for the fit
with no positivity conditions taken into account). Power law
corrections are taken in the simple form (suggested in \cite{q2})
$h_{1p}/Q^2$ for $g_{1}^{p}$ and $h_{1n}/Q^2$ for $g_{1}^{n}$ (and
corresponding combination for $g_{1}^{d}$). Fit A99 corresponds to the
fit to experimental data on polarized structure functions, B99
to $g_{1}$ but with $F_{2}$ taken from formula from \cite{SMCth} and C99 to the fit to spin
asymmetries where power law corrections are negligible. In the
Table 1 we also give $\chi^{2}$ for corresponding fits without power
law corrections. By comparing corresponding columns we see how the
introduction of these corrections reduces $\chi^{2}$. It is
seen from Table 1 that integrated parton densities do not differ
much for the three fits (It is also true in the case of fits
without power law corrections) however there is essential increase
in first two models relatively to the third. It seems that the
increase in $\chi^{2}$ of fits A99 and B99 in comparison with fit
C99 is connected with positivity conditions assumed for the parton
distributions.

To show the influence of positivity conditions we present also the
values for $\chi^{2}$ for the fits without positivity conditions
taken into account with power corrections  and without such
corrections. The values of integrated quark densities do not
change significantly (the strongest change is in gluon
contributions which become positive) in the fits without
positivity conditions comparing with those where positivity
conditions for parton densities were assumed. From Table 1 we see
that in the models without positivity conditions the effect of
taking into account power low corrections is rather small, i.e.
2.6 in the first ($h_{1p}=0.02\pm 0.01$, $h_{1n}=0.01\pm 0.04$)
and 2 in the second model ($h_{1p}=0.02\pm 0.02$, $h_{1n}=0.03\pm
0.04$) for 2 degrees of freedom. That means that in those fits the
parameters $h_{1p}$ and $h_{1n}$ are at the border of being
relevant. It is not very surprising that the fits with assumed
positivity conditions (that means with additional restrictions on
the form of the fitted parton densities) give $\chi^{2}$ higher
then fits where the positivity conditions are not assumed. On the
other hand we probably should not expect in NLO QCD very strong
violation of positivity conditions. The values for the integrated
parton densities given in Table 1 for the third model can be
compared with the values obtained using MRST 98 model:
\begin{eqnarray}
\Delta u& =& 0.77 ,  \nonumber \\ \Delta d&= &-0.59 ,\nonumber \\
\Delta s&= &-0.20, \nonumber \\ \Delta\Sigma &=&-0.02, \\
g_{A}&=&1.36 \nonumber \\ \Delta G&=&0.01 \nonumber \\
\chi^{2}&=&150.47.\nonumber
\end{eqnarray}
Our fit to spin asymmetries where 431 points were considered
(without averaging over $Q^{2}$) was discussed in \cite{BTpr} and similar
values for integrated parton densities were obtained.

In Table 2 we present in a similar way as in Table 1 results
obtained from the fit corresponding to the fit MRST 2001.

\vspace{0.5cm}
\hspace{5cm} Table 2

\vspace{0.2cm}
\hspace{0.5cm}
\begin{tabular}{||c|c|c|c|c|c|c||}
\hline \hline $fit$&$\Delta u$&$\Delta d$&$\Delta s$&$\Delta
\Sigma$&$g_A$&$\Delta G$\\ \hline \hline $A01$&$0.79
$&$-0.32$&$-0.04 $&$0.47$&$1.14$&$56.6$ \\ \hline $B01$&$0.81
$&$-0.35$&$-0.06 $&$0.40$&$1.16$&$42.6$ \\ \hline $C01$&$0.86
$&$-0.38$&$-0.05 $&$0.44$&$1.24$&$29.9$ \\ \hline \hline
$$&$h_{1p}$&$h_{1n}$&$\chi^2$&$\chi^2_0$&$\chi^2_{(np)}$&$\chi^2_{0(np)}$\\
\hline \hline $A01$&$0.06
$&$0.00$&$199.73$&$217.44$&$155.10$&$158.51$ \\ \hline $B01$&$0.05
$&$0.02$&$188.88$&$197.73$&$152.77$&$154.41$ \\ \hline
$C01$&$-0.0005 $&$-0.002$&$158.40$&$158.41$&$148.14$&$148.23$ \\
\hline \hline
\end{tabular}
\vspace {0.5cm}

The values of $\chi^{2}$ corresponding to the model with
positivity condition for parton densities in this case are much
higher then for the solution corresponding to MRST 99 presented in
Table 1. For the spin asymmetries one has $\chi^2$=158.41, which
have to be compared with 150.75. These solutions have more
singular behaviour for valence $u$ quark and very different gluon
behaviour for small $x$. Therefore  it is more difficult to fit
(with positivity conditions for quark densities) the experimental
data and $\chi^{2}$ for fits A01 and B01 and C01 is higher. On the
other hand the solutions without assuming positivity conditions
are not very different from the solutions corresponding to MRST 99
($\chi^{2}$ corresponding to spin asymmetries 148.23 is smaller
then 149.58). The reduction in $\chi^{2}$ coming from power law
corrections for solutions with assumed positivity conditions for
quark densities is significant (for B01 $h_{1p}=0.05\pm 0.02$,
$h_{1n}=0.02\pm 0.04$) but similarly to the situation presented in
Table 1 for the solutions without assuming positivity conditions
the changes in $\chi^{2}$ caused by these corrections are rather
small (for A01$(np)$ $h_{1p}=0.03\pm 0.02$, $h_{1n}=-0.01\pm 0.04$
and for B01$(np)$ $h_{1p}=0.02\pm 0.02$, $h_{1n}=0.01\pm 0.04$).
It is difficult to draw a conclusion that taking into account our
simple power law corrections of the form $h/Q^2$ is very important
for the fits. The integrated quark densities for different fits do
not differ much (like in Table 1 there are some changes for
integrated glouon distributions). In general taking into account
positivity conditions for parton densities, in spite of relatively
big changes in $\chi^{2}$, does not change significantly the
integrated parton densities.

There are changes when we compare solutions corresponding to MRST
98, MRST 99 and MRST 2001. From the eq.[6], Table 1 and Table 2 we
see that $\Delta u$ and $\Delta d$ increase, $\Delta s$ decrease
and as consequence $\Delta \Sigma$ increase. The values of $\Delta
\Sigma$ are not very different from $g_{8}=0.58$. We have the
situation when the solutions give different integrated parton
densities  and the corresponding $\chi^{2}$ values are very close
(especially when we take solutions corresponding the case when we
do not assume positivity conditions for parton densities). There
is a dependence on the form of assumed parton densities but that
does not influence strongly $\chi^{2}$ values. It is tempting to
choose (in spite of higher $\chi^{2}$ value when positivity
condition for quark densities are assumed) solution corresponding
to MRST 2001 (recent fit to unpolarized data) where integrated
parton densities give relatively high value of $\Delta \Sigma$ and
small value of $\Delta s$ (or at least consider it seriously).

We have compared polarized parton distributions, corresponding to
MRST 99 and MRST 2001 determination, using three different methods
of obtaining polarized parton densities. Simplest power law
correction corresponding to higher twists have been taken into
account. As expected these power law corrections are negligible in
the fits to spin asymmetries. At the first sight they seem to be
important in the first two models. For comparison we have also
considered the solutions where positivity conditions for parton
densities have not been assumed. We observe strong increase in
$\chi^{2}$ in models A99, A01 and B99, B01 (fits to $g_{1}$)
connected with positivity conditions. Reduction in $\chi^{2}$ by
taking into account power law corrections is significant in the
models where positivity conditions were assumed and rather
marginal in models without positivity conditions for parton
densities. Hence, it seems that the determination of power law
corrections is not very reliable. There are some differences in
integrated parton densities corresponding to the models MRST98,
MRST 99 and MRST 2001. Within the definite model three different
methods of fitting give very similar results for integrated quark
densities (but not very much for a gluon one) also in the case
when positivity conditions for parton densities are not assumed.
The latest model (C01) even if not by $\chi^{2}$ (with slightly
higher $\chi^{2}$ value with positivity conditions for parton
densities assumed) values is preferred by interpretation reasons.

\end{document}